\documentstyle[twocolumn,aps,psfig,epsf]{revtex}

\begin{document}
\twocolumn[\hsize\textwidth\columnwidth\hsize\csname 
 @twocolumnfalse\endcsname
\title{Creep of superconducting vortices in the limit of vanishing temperature:
A fingerprint of off-equilibrium dynamics.}
\author{Mario Nicodemi$^{a,b}$ and Henrik Jeldtoft Jensen$^{a}$}
\pagestyle{myheadings}
\address{$^{a}$ Department of Mathematics, Imperial College, 180 Queen's Gate, 
London SW7 2BZ, UK \\ 
$^b$ Universit\'a di Napoli ``Federico II'',
Dip. Scienze Fisiche, INFM and INFN, Via Cintia, 80126 Napoli, Italy}

\maketitle
\date{\today}
\begin{abstract}
We theoretically study the creep of vortex matter 
%the inhomogeneous vortex density profile (the Bean state) 
in superconductors. The low temperatures 
experimental phenomenology, 
previously interpreted in terms of ``quantum tunnelling'' of vortices, 
is reproduced by Monte Carlo simulations of a purely ``classical'' 
vortex model. We demonstrate that a non-zero creep rate in the limit
of vanishing temperature is to be expected in systems with slow relaxations 
as a consequence of their off-equilibrium evolution in a complex free energy 
landscape. 
\end{abstract}
\vskip1pc

]

%\section{Introduction}
There exist an abundance of experimental evidence that the relaxation rate of 
the magnetisation in type II superconductors (ranging from conventional, 
to organic, heavy fermions and 
a variety of $H-T_c$ superconductors \cite{mota,fruchter,stein})
does not vanish as the 
temperature $T$ is lowered towards zero (see eg. \cite{mota,fruchter,stein}). 
This may seem surprising if one assumes that the mechanism allowing the 
magnetisation to relax is thermal activation over a characteristic energy 
barrier $\Delta E$. Namely, when $T\ll \Delta E$ the relaxation rate should 
vanish since the Arrhenius factor for thermal activation, $\exp(-\Delta E/T)$, 
goes to zero. The question then arises how relaxation can 
%continue to 
take place at a 
finite rate while the thermal activation factor exponentially approaches zero. 

A number of researchers have suggested that such a phenomenon is caused 
by quantum tunnelling of vortices through the barriers in the random pinning 
potential (for a theoretical review see \cite{blatter}). The above ``quantum'' 
explanation is very intriguing and in good agreement with 
some experimental results in compounds such as YBCO \cite{mota,fruchter} 
or BSCCO \cite{aupke}. Also other even more exotic materials, such as 
organic superconductors \cite{mota_org}, give good correspondences. 
However, in other non conventional systems such as heavy fermion 
superconductors, the theory of quantum creep is totally unable to 
describe the observed low $T$ relaxation \cite{mota_hf}. 
In fact, strong discrepancies are found in many other 
systems ranging from PCeCO crystals to YBCO/PBCO multilayers or 
YBCO and BSCCO films and crystals \cite{hoekstra}. One problem is 
\cite{hoekstra} that the length of the tunnelling vortex segment $L_c(0)$ 
needed to fit the creep rate data, $S$, can be orders of magnitude larger 
that the one 
theoretically predicted by quantum creep theory \cite{blatter}.
%\cite{note1}. 
Also the experimental temperature dependence 
of the creep rate, $S$, is often very different from the one predicted 
by quantum theory 
%(typically linear instead of quadratic, 
%see for instance \cite{aupke}). 
(see \cite{aupke}).

The above contradictory  results 
%and the idea that some general simple 
%mechanism might still be present in all the above materials naturally 
suggests to look for additional descriptions of the anomalous low $T$ magnetic 
relaxation. It is worth to stress that the observation of a non-vanishing 
constant creep rate in the limit $T\rightarrow 0$ is found under very general 
circumstances: 
it does not crucially depend on the thickness of the sample 
\cite{hoekstra} (i.e. on its dimensionality), 
nor on whether the pinning is caused by columnar defects or random point 
pins \cite{nowak}. 
Thus, the mechanism behind the low temperature creep seems to be of a 
fundamental and basic nature. 

We demonstrate below that also in a ``classical'' system 
(i.e., not ``quantum'') 
logarithmically slow glassy dynamics can naturally persist even at vanishing 
temperatures and can lead to the experimentally observed phenomenology. 
This is possible because the low $T$ off-equilibrium dynamics consists 
of searching, among a very large number, for a few ``downhill'' or ``flat'' 
directions in the free energy landscape. 
The number of these directions decreases as relaxation proceeds 
though there always remain some. They can be found only by collective 
cooperative rearrangements of the system, resulting in a slowing down of 
relaxation \cite{EAN}. 
%We discuss this scenario in terms of a simple model which was 
%shown to reproduce a broad range of experimental facts of vortex dynamics.

{\it The model --} 
We study a statistical mechanics model for vortex matter 
called a Restricted Occupancy Model (ROM) \cite{nicodemi1}. 
In the limit of zero temperature and infinite upper critical field, 
it reduces to a cellular automaton 
introduced in \cite{bassler1} to study vortex avalanches. 
We use Monte Carlo dynamics which enabled the ROM to depict a unified 
picture of creep and transport 
phenomena in vortex physics, ranging from magnetisation loops with 
``anomalous'' second peak, logarithmic relaxation, Bean profiles, 
to history dependent behaviours in vortex flow and I-V characteristics, 
to the reentrant nature of the equilibrium phase diagram \cite{nicodemi1}. 
The model also predicts the existence of a ``glassy region'' 
at low temperature with strong ``aging'' effects \cite{nicodemi1}. 

Here we use the ROM to study the magnetic relaxation rate, 
$S$, in the very low temperature limit. Interestingly, 
the model reproduces the experimental ``anomalous'' relaxation 
and the observed behaviour of $S$ \cite{mota,fruchter,stein,hoekstra}. 

%The model is an attempt to identify the essential degrees of freedom 
%and the effective dynamics responsible for the experimental observations 
%\cite{nicodemi1}.
The ROM model is described in full details in Ref.\cite{nicodemi1}.
A system of straight parallel vortex lines is coarse grained in the 
$xy$-plane by introducing a square grid of lattice spacing $l_0$ 
of the order of the London penetration length, $\lambda$ \cite{nicodemi1}. 
The number of vortices on the $i$-th coarse grained plaquette 
is denoted by $n_i$. The occupancy of 
each plaquette is a number larger than zero and, importantly, smaller
than $N_{c2}=B_{c2}l_0^2/\phi_0$, 
where $B_{c2}$ is the upper critical magnetic field 
and $\phi_0=hc/2e$ is the magnetic flux quantum. 
The ROM model is thus defined by the following coarse grained 
vortex interaction Hamiltonian \cite{nicodemi1}: 
%\begin{equation}
${\cal H}= \frac{1}{2} \sum_{ij} n_i A_{ij} n_j 
-\frac{1}{2} \sum_i A_{ii} n_i - \sum_i A^p_i n_i$.
%\label{H}
%\end{equation}
The first two terms describe the repulsion between the vortices and their 
self energy. On-site and nearest neighbour interactions are included, i.e.,  
the interaction between vortex lines with a separation 
greater than the London screening length 
(which by definition is close to $l_0$) is ignored. 
We choose $A_{ii} = A_0=1$; $A_{ij}=A_1$ if $i$ and $j$ are nearest neighbours 
and $A_{ij}=0$ otherwise. The last term in ${\cal H}$ describes a 
delta-distributed random pinning $P(A^p)=(1-p)\delta(A^p)+p\delta(A^p-A^p_0)$. 
Interestingly, the general scenario of creep phenomena we describe below 
does not depend on the details of pinning in the system
(a fact in correspondence with experimental results 
\cite{hoekstra}). In our model $A_0$ sets the energy scale. Below we choose 
$A_0=1.0$; $A_0^p=0.3$; $N_{c2}=27$; $p=1/2$; 
$\kappa^*\equiv A_1/A_0\in[0.26,0.3]$. 
%The ROM model is explained in all its details in Ref.\cite{nicodemi1}, 
%here we only notice that 
%a theoretical justification of its Hamiltonian consists in the 
%observation that the repulsive interaction between vortices of spatial 
%separation less than $\lambda$ is included and that the existence of the 
%upper critical field is represented by the constraint $n_i\leq N_{c2}$. 

The relaxation of the model is simulated by use of Monte Carlo dynamics 
on a square lattice in presence of a thermal bath of temperature $T$. 
The system is periodic in the $y$-direction. The two edges parallel to the 
$y$-direction are in contact with a vortex reservoir. 
%, described by the above Hamiltonian without pinning. 
Particles can enter and exit the system only through the reservoir, 
which plays the role of the external magnetic 
field. Hence the reservoir density, $N_{ext}$, is  used as the external 
control parameter. We perform the following zero field 
cooled experiment: at a low temperature $T$ we increase at constant rate 
$\gamma= \Delta N_0/\tau$ the  reservoir 
density from zero to a working value $N_{ext}$.
We keep $N_{ext}$ fixed while we 
monitor the time dependence of the magnetisation $M=N_{in}-N_{ext}$. 
Here $N_{in}=\sum_in_i/L^2$ is the vortex density inside the system (of 
size $L^2$ \cite{nota_L}). 
Time is measured in units of single attempted updates per degrees of 
freedom of the lattice (see Ref. \cite{Binder}). 
The data presented below are averaged over 128 realizations of the pinning
background.

In particular, we investigate the creep rate 
\begin{equation} 
S=\left|{\partial \ln (M)\over \partial \ln(t)}\right| 
\end{equation} 
as function of $T$, $N_{ext}$ and $\gamma$. 
In typical experiments the nature of the $t$ dependence of $M$ is such that  
$S$ decreases in time. So usually, one deals with an average creep rate, 
$S_a$, in some given temporal window
\cite{mota,fruchter,stein,blatter,aupke,mota_org,mota_hf}. 
As shown in the upper inset of Fig.\ref{S_0}, 
in our model dynamics $M(t)$ at low temperatures behaves according to
the known logarithmic interpolation formula 
(see Ref.\cite{nicodemi1}) 
found in experiments \cite{blatter}, namely:
$M(t)-M(0)\simeq \Delta M_{\infty}
\{1-[1+
\frac{\mu T}{U_c}\ln(\frac{t+t_0}{t_0})]^{-1/\mu}\}$. 
%\begin{equation}
%M(t)-M(0)\simeq \Delta M_{\infty}
%\left\{1-\left[1+
%\frac{\mu T}{U_c}\ln\left(\frac{t+t_0}{t_0}\right)\right]^{-1/\mu}\right\}. 
%\label{mag_log}
%\end{equation}
Here, $M(0)$ is the magnetisation at the time of preparation of the sample 
($t=0$), $\Delta M_{\infty}$ its overall variation, the exponent 
$\mu$ is consistent with 1, $\frac{\mu T}{U_c}$ and $t_0$ are fit parameters 
\cite{blatter}. 

Consistently, we define $S_a$ as the average 
value of $S$ in the last time decade of our measures 
(i.e., for $t\in[10^5,10^6]$).
%\begin{equation}
%S_a\equiv {1\over (t_1-t_0)} \int_{t_0}^{t_1} S(t) ~ dt
%\end{equation}
%with $t_0=10^5$ and $t_1=10^6$. 
The present choice, analogous to those made for experimental data, 
is the most natural one 
and the general results presented below do not depend on it.
%It is worth noting that experimentalists are forced to make similar choices
%concerning the specific way $S_a$ is measured \cite{aupke}. 

For the reasons explained in the introduction, a very important physical 
quantity is the distribution, $P(\Delta E)$, of the energy barriers, 
$\Delta E$, that vortices segments  meet during their motion. 
Since at low $T$ the system 
is typically off-equilibrium, $P(\Delta E)$ is itself a (logarithmically 
slow) function of $t$ and 
%Consistently with the above definition of $S_a$, 
we consider its 
%energy barrier  distribution 
average over the last time decade of our measurements.

\begin{figure}[ht]
\vspace{-1.5cm}
\centerline{
\hspace{-2.5cm}\psfig{figure=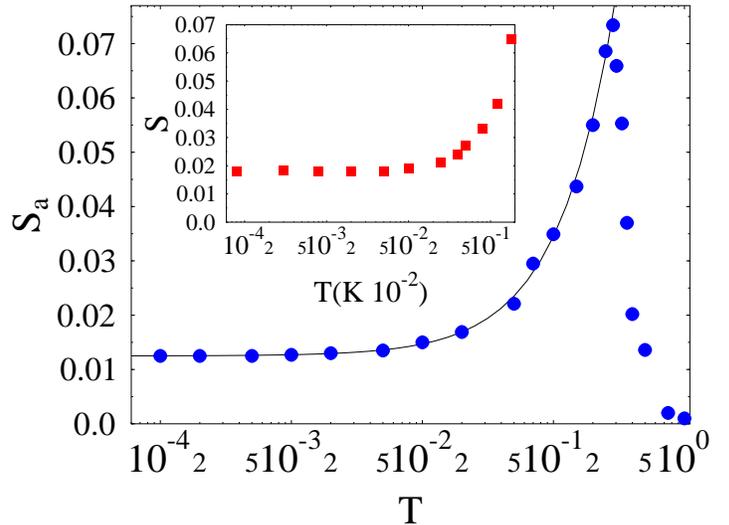,width=8cm,angle=-90} 
}
\vspace{-1.6cm}
\caption{{\bf Main frame} The creep rate, $S_a$, in the ROM model 
($\kappa^*=0.28$, $\gamma=10^{-3}$) for $N_{ext}=10$ as a function of 
the temperature, $T$, in units of $A_0$. The error bars are of the size of 
symbols. The superimposed line is a linear fit. 
{\bf Inset} The creep rate measured in BSCCO single crystal from 
Aupke et al. [5]  for an external field of 880 Oe.} 
\label{S_T}
\end{figure}

{\it Results --}
When the temperature is very low, the model exhibits the same kind of 
``anomalous'' creep found in the experiments on superconductors. 
In Fig.\ref{S_T}, we plot the creep rate, $S_a$, 
as a function of $T$ in a broad temperature range. 
For comparison we present equivalent experimental measurements in a BSCCO 
single crystal (from Ref.\cite{aupke}) as inset. 
The numerical values found for $S_a$ at low $T$ in our model and in real 
samples are interestingly very similar. 
The temperature scales of the simulations and of real 
experiments can be compared by considering that 
the ratio $T/A_0$ in our model is of the same order of magnitude as $T/T_c$ 
in a real superconductor. This is seen from a comparison of the $(T,N_{ext})$ 
equilibrium phase diagram of our model with the equilibrium 
temperature-magnetic field, $(T,H)$, phase diagram of, say, a BSCCO 
superconductor (see Ref.\cite{nicodemi1}). 

%Apparently, the simulations and the measurements on superconductors exhibit 
%very similar behaviour. 
In both experiments and simulations, 
$S_a$ approaches a finite value, $S_a^0$, 
when $T\rightarrow 0$. In particular, we find that a linear fit of 
$S_a(T)$ in the low $T$ regime is very satisfactory (see Fig.\ref{S_T}): 
\begin{equation}
S_a(T)=S_a^0+\sigma T
\end{equation}
where both $S_a^0$ and $\sigma$ are a function of the applied field 
$N_{ext}$. 
%For a given value of $N_{ext}$, the above fit can be improved 
%by including higher powers of $T$. 
We also note that in the present model 
$S_a(T)$ is non monotonous in $T$: in the higher $T$ region it 
starts decreasing. This is also a known experimental fact 
\cite{mota_org,Yeshurun}, we discuss it later on. 
The maximum in $S_a(T)$ is just above a characteristic crossover ``glassy'' 
temperature, $T_g$, defined in \cite{nicodemi1}. 

\begin{figure}[ht]
\vspace{-1.5cm}
\centerline{
\hspace{-2.5cm}\psfig{figure=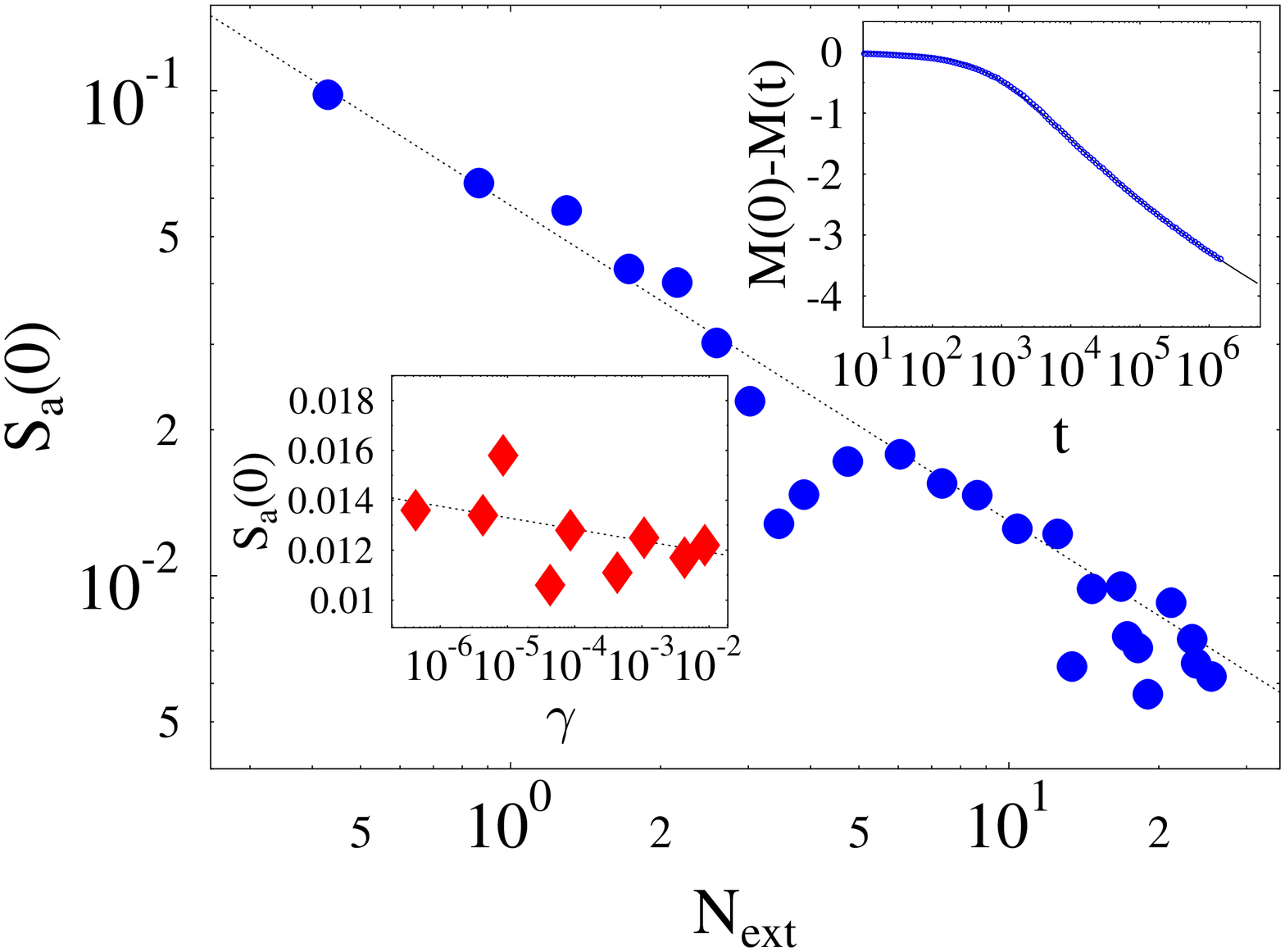,width=10cm,angle=-90} 
}
\vspace{-2.4cm}
\caption{{\bf Main frame} The zero temperature limit of the creep rate, $S_a$, 
in the ROM model as a function of the applied field $N_{ext}$ 
(for $T=10^{-4}$ and $\gamma=10^{-3}$). 
The superimposed curve is a power law to guide the eye.  
{\bf Inset top:} 
The relaxation of the magnetisation, $M(t)$, in the model 
($\kappa^*=0.28$, $\gamma=10^{-3}$) for $N_{ext}=10$ and $T=0.25$ 
as a function of time. 
The continuous line is the logarithmic fit of the text. 
{\bf Inset bottom:} 
$S_a$ is here plotted as a function of the sweep rate, $\gamma$, 
of the external field for $N_{ext}=10$ and $T=10^{-4}$. A very weak 
dependence is found.} 
\label{S_0}
\end{figure}

%It is interesting that a very similar numerical 
%value for $S_a^0$ is found in both model and experimental data. 
%Experimentally, for many different systems, $S_a^0$ values are typically 
%around the interval $10^{-2}-10^{-1}$
%\cite{mota,fruchter,stein,blatter,aupke,mota_org,mota_hf}. 
In our model 
%(for the examined value $\kappa^*=0.28$) 
by varying the applied field
we find a range of values for $S_a^0$ very similar to experimental ones 
\cite{mota,fruchter,stein,blatter,aupke,mota_org,mota_hf}
(see Fig.\ref{S_0}). In particular, 
$S_a^0$ seems to decrease on average by increasing the field $N_{ext}$. 
The overall behaviour can be approximately interpolated with a power law:
$S_a^0(N_{ext})\simeq (N_{ext}/N_0)^{-x}$, where, for $\kappa^*=0.28$, 
$N_0\simeq 0.01$ and $x\simeq 0.6$. As shown in Fig.\ref{S_0}, 
the presence of a small exponent $x$ implies that 
%$S_a^0$ is slowly varying with $N_{ext}$: 
sensible variations in $S_0^a$ can be seen only 
by changing $N_{ext}$ of orders of magnitude. 
Note that the dips in the $S_a(0)$ versus $N_{ext}$ data in 
Fig. \ref{S_0} at certain values of $N_{ext}$ (namely around 3, 13, and 18) 
are statistically significant. 
They are related to 
 the low field order-disorder transition, the 
2nd peak transition and the reentrant high field order-disorder 
transition respectively \cite{nicodemi1}. 

In the lower inset of Fig.\ref{S_0}, we show that $S_a$ is essentially 
independent of the ramping rate, $\gamma$ (the values shown are for 
$N_{ext}=10$ and $T=10^{-4}$). This is an other typical experimental 
observation \cite{mota}. However, a very small decrease of $S_a$ with 
increasing $\gamma$ cannot be excluded: we show a fit to the form
$S_a^0\simeq S_1+s_2\ln(\gamma)$, with $S_1=0.11$ and $s_2=-2\cdot 10^{-4}$. 
The fact that $S_a^0$ is practically independent on $\gamma$, far 
from being a proof of the presence of equilibrium in the system, 
is due to the fact that at very low temperatures 
the characteristic equilibration time, $\tau_{eq}$, is enormous 
(see Ref.\cite{nicodemi1}). So whenever the driving rate, $\gamma$, 
is much larger than $\tau_{eq}^{-1}$  
the off-equilibrium state and dynamics of the system are essentially independent of $\gamma$. Stronger $\gamma$ effects 
have to be expected when $\gamma$ gets closer to $\tau_{eq}^{-1}$. 
In fact, it is experimentally well known that at higher temperatures 
the systems exhibit strong $\gamma$ dependent 
``memory'' effects \cite{nicodemi1,Yeshurun,memoryeffects}, 
the signature of off-equilibrium dynamics. 
Actually, in the present model at low $T$, it is possible to show that 
$\tau_{eq}(T)$ diverges exponentially \cite{nicodemi1}: 
$\tau_{eq}(T)\sim \exp\left({E_0\over T}\right)$. 
In that region, the typical observation time windows, $t_{obs}$, are 
such that $t_{obs}/\tau_{eq}\ll 1$, and 
the system is in the early stage of its off-equilibrium relaxation 
from its initial state. 
This is schematically the origin of the flattening of $S_a$ at very low $T$. 
Notice that, if one could observe the system for an exponentially long time, 
i.e., if $t_{obs}/\tau_{eq}\gg 1$, then the creep rate, $S_a$, would 
indeed go to zero.

\begin{figure}[ht]
\vspace{-1.5cm}
\centerline{
\hspace{-2.5cm}\psfig{figure=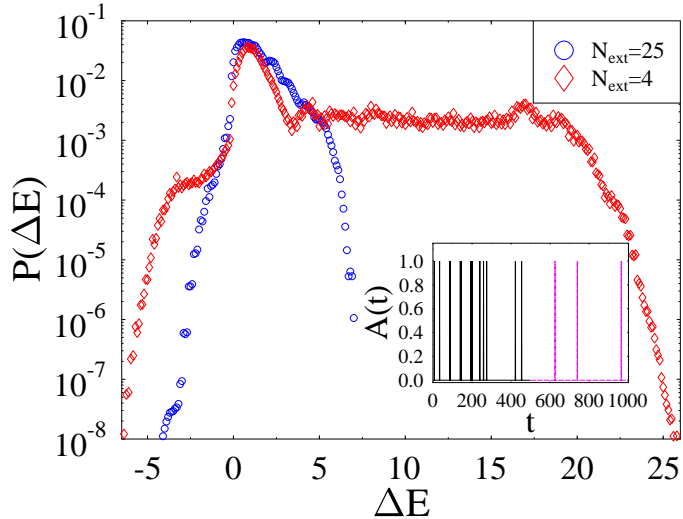,width=8cm,angle=-90}}
\vspace{-1.6cm}
\caption{The energy barrier distribution, $P(\Delta E)$, recorded 
at $T=10^{-4}$ for $N_{ext}=25,4$ (resp. circles,diamonds). 
Notice that $P(\Delta E)$ extends down to negative values.
{\bf Inset } We plot the function $A(t)$ defined in the text to monitor 
the activity during relaxation. In the interval $[0,500]$ we show 
$A(t)$ recorded at the beginning of the run and in $[500,1000]$ 
$A(t)$ in the last steps of the same run.} 
\label{P_E}
\end{figure}

The above scenario is clarified by the analysis of the energy barrier 
distribution function, $P(\Delta E)$, recorded during the system evolution at 
very low $T$. Such a quantity also clearly shows the simple 
mechanical origin of the anomalous creep found at very low temperature 
in the present model. The function $P(\Delta E)$ (where $\Delta E$ is in units of $A_0$), 
recorded at $T=10^{-4}$, is plotted in Fig.\ref{P_E} 
for two values of the applied field, $N_{ext}$. We always find that
 $P(\Delta E)$ has support 
also on the negative axis. This is the mark of the off-equilibrium nature 
of the evolution on the observed time scales.
%, because, by definition,
%in mechanical equilibrium only non-negative energy barriers should be found. 
The presence of a $P(\Delta E)$ which extends down to negative 
values also explains the presence of the recorded relaxation at 
low $T$: in the configuration space the system can still find 
directions where no positive barriers have to be crossed.
The insert in Fig. \ref{P_E} clarifies the mechanism behind the relaxation.
Here, we plot the signal $A(t)$ defined, for each single Monte Carlo (MC) 
step $t$, in the following way: $A(t) = 0$ if the MC trial is 
rejected; $A(t)=1$ if the trial is accepted and the energy reduced, 
i.e. $\Delta E\leq 0$; and finally $A(t)=2$ when a trial is 
accepted with $\Delta E>0$. We plot two sequences  of trials.
One for $0\leq t\leq 500$ was measured at the early stage of the relaxation,
the second sequence, placed at the interval $500\leq t\leq 1000$, is measured 
at the late stage of the relaxation. Most trials are rejected ($A(t)=0$) and 
only once in a while the system does manage to find a route pointing
downhill in the energy landscape. $A(t)=2$ never 
occurs. As time proceeds fewer and fewer ``negative channels'' are available 
to the relaxation and a decrease in the density of the spikes 
in $A(t)$ is observed. 

As the temperature is increased thermal activation over positive
energy barriers will become possible as the  Arrhenius factor 
$\exp(-\Delta E/T)$
assumes a non-vanishing value for an appreciable range of barrier
values $\Delta E<T$. 
When this happens the relaxation will occur sufficiently fast
to allow one, within the experimental time window, to closely approach 
the equilibrium configurations where the vortex density profile is more or less
flat and relaxation ceases, hence $S_a$ goes down, as seen in experiments
\cite{mota_org,Yeshurun} and in Fig. 1.

Finally, we stress that slow off-equilibrium relaxations 
at very low temperatures 
%(below the ``freezing'' point) 
are also observed in glass forming liquids \cite{EAN}. 
In that cases too, no activation 
over barriers occurs and the system simply wanders in its very high 
dimensional phase space through the few channels where no energy increase 
is required. 

{\it Discussion} - We have above demonstrated that the phenomenological 
behaviour of the creep rate at low temperatures can be understood in terms 
of the off-equilibrium nature of the inhomogeneous vortex density profile 
produced in magnetic creep experiments. 
In fact, cooperative mechanical rearrangements, possible even at very low $T$ 
(where thermal activation over positive barriers can be negligible), 
dominate the phenomenon \cite{EAN}. In this perspective, 
it is very important to stress that the system's equilibration time at very 
low temperature is much larger than any 
experimentally accessible time window \cite{nicodemi1}. 
Accordingly, we can say that experimental findings do not enforce the 
interpretation in terms of macroscopic quantum tunnelling of vortices. 

Relaxation due to quantum tunnelling might be present along with 
the mechanical relaxation discussed above. 
It is then important to ask how compelling the quantum tunnelling 
interpretation is. The theory 
of quantum tunnelling assumes a London picture and treats the 
position of the vortex core as the variable that is able to tunnel. 
It is not entirely clear if this is the 
right level at which to introduce quantum fluctuations, but, more importantly, 
the quantum tunnelling description 
assume the existence of a characteristic energy barrier \cite{blatter},
not a time dependent distribution of barrier heights as typically found 
in many-body systems relaxing off-equilibrium. 
%The scale of such a barrier is estimated in 
%from single pining or from collective pinning arguments, but in any case 
%a ``mean field-like'' approach, 
%where the fluctuations in the sizes of the activation 
%barriers, arising from many-body effects in vortex-vortex or vortex-pin 
%interactions, are neglected. 
Finally, the quantum tunnelling 
description also tacitly assumes the existence of a static equilibrium 
state in which barriers are always positive. As we have clearly shown 
above, this is typically not the case and a dynamical approach is more 
appropriate. 

The present scenario, where off-equilibrium phenomena 
dominate the anomalous low $T$ creep, could be experimentally verified by the 
discovery of ``aging'' phenomena \cite{nicodemi1} 
like those recently observed in 
%vortex physics 
\cite{Papad,andrei}.

Work supported by EPSRC, INFM-PRA(HOP)/PCI.


\begin{thebibliography}{40}
\bibitem{nota_L} We take $L=32$, but we checked our results in the 
size range $L=8,16,32,64,128$.
\end{thebibliography}
\end{document}